\documentclass[a4paper,11pt]{article}
\usepackage{pos}
\usepackage{braket}
\usepackage{xcolor}
\usepackage[normalem]{ulem}
\usepackage{enumitem}
\usepackage{natbib}

\title{An update on QCD+QED simulations with C$^\star$ boundary conditions}

\author[a]{Lucius Bushnaq}
\author[b]{Isabel Campos}
\author[c]{Marco Catillo}
\author[d]{Alessandro Cotellucci}
\author[ef]{Madeleine Dale}
\author[a]{Patrick Fritzsch}
\author*[dg]{Jens L\"ucke}
\author[c]{Marina Krsti\'c Marinkovi\'c}
\author[dg]{Agostino Patella}
\author[ef]{Nazario Tantalo}

\affiliation[a]{School of Mathematics, Trinity College Dublin, Dublin 2, Ireland}
\affiliation[b]{Instituto de F\'isica de Cantabria \& IFCA-CSIC, Avda. de Los Castros s/n, 39005 Santander, Spain}
\affiliation[c]{Institut f\"ur Theoretische Physik, ETH Z\"urich,  Wolfgang-Pauli-Str. 27, 8093 Zürich, Switzerland}
\affiliation[d]{Humboldt Universit{\"a}t zu Berlin, Institut f{\"u}r Physik \& IRIS Adlershof, Zum Gro{\ss}en Windkanal 6, 12489 Berlin, Germany}
\affiliation[e]{Universit\`a di Roma Tor Vergata, Dip. di Fisica, Via della Ricerca Scientifica 1, 00133 Rome, Italy}
\affiliation[f]{INFN, Sezione di Tor Vergata, Via della Ricerca Scientifica 1, 00133 Rome, Italy}
\affiliation[g]{DESY, Platanenallee 6, D-15738 Zeuthen, Germany}

\emailAdd{jens.luecke@hu-berlin.de}
\emailAdd{agostino.patella@physik.hu-berlin.de}
\emailAdd{fritzscp@tcd.ie}
\emailAdd{isabel.campos@csic.es}
\emailAdd{bushnaql@tcd.ie}
\emailAdd{m.dale@stimulate-ejd.eu}
\emailAdd{nazario.tantalo@roma2.infn.it}
\emailAdd{mcatillo@ethz.ch}
\emailAdd{marinama@phys.ethz.ch}
\emailAdd{alessandro.cotellucci@physik.hu-berlin.de}

\abstract{

\begin{center}
\large
\includegraphics[width=.11\textwidth]{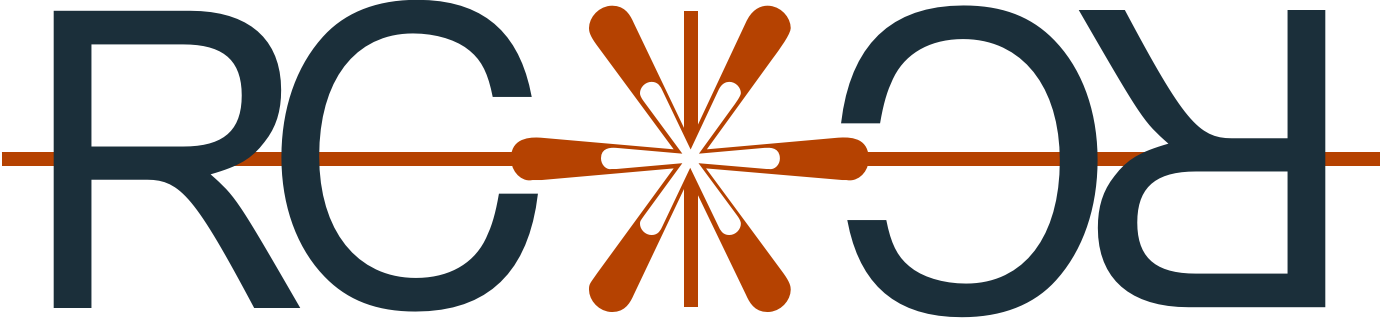} \hspace{1ex} collaboration
\end{center}

\bigskip

We present two novelties in our analysis of fully dynamical QCD+QED ensembles with \cstar boundary conditions. The first one is the explicit computation of the sign of the Pfaffian. We present an algorithm that provides a significant speedup compared to traditional methods. The second one is a reweighting of the mass in the context of the RHMC. We have tested the techniques on both pure QCD and QCD+QED ensembles with pions at $m_{\pi^\pm}\approx400$ MeV, a lattice spacing of $a\approx0.05$ fm, a fine-structure constant of $\alpha_{\mathrm{R}}=0$ and $0.04$.}

\FullConference{%
 The 38th International Symposium on Lattice Field Theory, LATTICE2021
  26th-30th July, 2021
  Zoom/Gather@Massachusetts Institute of Technology
}

\newcommand{\cstar}{$\mathrm{C}^{\star}\,$}

\newcommand{\agorm}[1]{}
\renewcommand{\agorm}[1]{{\color{red}\sout{#1}}}

\bibliography{lattice.bib}

\begin{document}
\maketitle


\section{Introduction}

We present an update on a long-term research program aiming at calculating isospin-breaking and QED radiative corrections in hadronic quantities, with C$^\star$ boundary conditions~\cite{Kronfeld:1990qu,Kronfeld:1992ae,Wiese:1991ku,Polley:1993bn} and fully-dynamical QCD+QED simulations.
C$^\star$ boundary conditions allow for a local and gauge-invariant formulation of QED in finite volume and in the charged sector of the theory~\cite{Lucini:2015hfa,Patella:2017fgk,Hansen:2018zre}. In particular, two ensembles were generated at the values of the fine-structure constant $\alpha_R=0.04$ and $\alpha_R=0$. A value of $\alpha_R$ larger than the physical one has been chosen to amplify QED corrections.

The open-source \texttt{openQ*D-1.1} code~\cite{openQxD-csic,Campos:2019kgw} was used to generate all gauge configurations presented in this work. This code has been developed by the RC$^\star$ collaboration. It is an extension of the \texttt{openQCD-1.6} code~\cite{openQCD} for QCD.

In this proceedings we will focus on two novelties in our analysis: the calculation of the sign of the Pfaffian of the Dirac operator (section \ref{sec:sign}), and a particular implementation of the mass reweighting in the context of the RHMC (section \ref{sec:mrw}).


\section{Simulation setup}

So far we have generated two $N_f=3+1$ QCD ensembles and two $N_f=1+2+1$ QCD+QED ensembles. We used the L\"uscher-Weisz action for the SU(3) field with $\beta=3.24$, the Wilson action for the U(1) field with $\alpha_0=0.05$ (for the QCD+QED ensembles), and $O(a)$-improved Wilson fermions. For the QCD ensembles, we used the value of $c_{\mathrm{sw}}$ determined non-perturbatively in~\cite{Fritzsch:2018kjg}. For the QCD+QED ensembles, in lack for a better option, we used the same value of $c_{\mathrm{sw}}$ for the SU(3) SW term, and $c_{\mathrm{sw}}=1$ for the U(1) SW term (see table~\ref{tab:parameters}).  We employ C$^\star$ boundary conditions in space and periodic boundary
conditions in time for all our ensembles. We have verified that we are free from the problem of topological freezing in all our ensembles, which justifies the use of periodic boundary
conditions in time.

Following \cite{Hollwieser:2019kuc}, we determine the lattice spacing from the auxiliary observable $t_0$, by using the central value of the CLS determination $(8t_0)^{1/2} = 0.415 \text{ fm}$ \cite{Bruno:2016plf}. This has been taken only as an indicative value, keeping in mind that it contains an $O(\alpha_0)$ ambiguity which can be resolved only when the scale is set with a physical observable, e.g. the mass of the $\Omega$ baryon. We obtain $a\simeq 0.054\text{ fm}$ for the QCD ensembles, and a marginally lower value for the QCD+QED ensembles (see table~\ref{tab:summary}).

We define the renormalized fine-structure constant $\alpha_\text{R}$ as
\begin{gather}
   \alpha_\text{R} = 
   \mathcal{N}^{-1}
   t_0^2
   \braket{E_{\mathrm{U(1)}}(t_0)}
\end{gather}
where $E_{\mathrm{U(1)}}(t)$ is the clover discretization of the U(1) action density calculated in terms of the gauge field at positive flow time $t$. The normalization $\mathcal{N}$ is chosen such that $\alpha_\text{R} = \alpha_0 + \mathcal{O}\left(\alpha_0^2\right)$. Our choice $\alpha_0=0.05$ for the QCD+QED ensembles, corresponds to an unphysically large value $\alpha_\text{R} \simeq 0.04 \simeq 5.5 \alpha_\text{R}^\text{phys}$.

In the QCD case, we have simulated the SU(3) symmetric point, i.e. $m_u = m_d = m_s \simeq (m_u+m_d+m_s)^\text{phys}/3$. In the QCD+QED case,we have chosen to work at the U-symmetric point, i.e. $m_d=m_s$, and we have chosen $m_u$ in such a way that the strong isospin-breaking effects are rescaled with the same factor as the QED isospin-breaking effects. The \textit{lines of constant physics} are determined by keeping the following quantities 
\begin{align}
   \phi_0 &= 8 t_0 ( m^2_{K^\pm} - m^2_{\pi^\pm} )
   \ , &&
   \phi_1 = 8 t_0 ( m^2_{\pi^\pm} + m^2_{K^\pm} + m^2_{K^0} )
   \ , \nonumber\\
   \phi_2 &= 8 t_0 ( m^2_{K^0} - m^2_{K^\pm} ) \alpha_\text{R}^{-1}
   \ , &&
   \phi_3 = \sqrt{8 t_0} ( m_{D_s} + m_{D^0}+ m_{D^\pm} )
   \ ,
\end{align}
constant as $\alpha$ is varied. While these quantities can be determined quite accurately from
lattice simulations, their real-world value is unknown, since $t_0$ cannot be
measured experimentally. In practice one needs to simulate different lines of
constant physics, and then interpolate/extrapolate to the
real-world one by setting the scale with a physical observable. The aim of this project is to simulate on the U-symmetric line of constant
physics defined by
\begin{gather}
   \phi_0 = 0 
   \ , \quad
   \phi_1 
   =
   2.13
   \simeq \phi_1^\text{phys}
   \ , \quad
   \phi_2 = 2.37 \simeq \phi_2^\text{phys}
   \ , \quad
   \phi_3 = 12.1 \simeq \phi_3^\text{phys}
   \ ,
\end{gather}
which, for $\alpha=0$, corresponds to the QCD SU(3)-symmetric point. In this setup the $\pi^\pm$ is heavier than the real-world one,
making simulations easier. In the context of QCD+QED, a similar strategy has been used e.g. in~\cite{Horsley:2015eaa}. As routinely done in QCD (and more so in the past),
one wants to start from heavier pions and then to approach the physical pion
mass in steps.

\begin{figure}
    \centering
    \includegraphics[width=.85\textwidth]{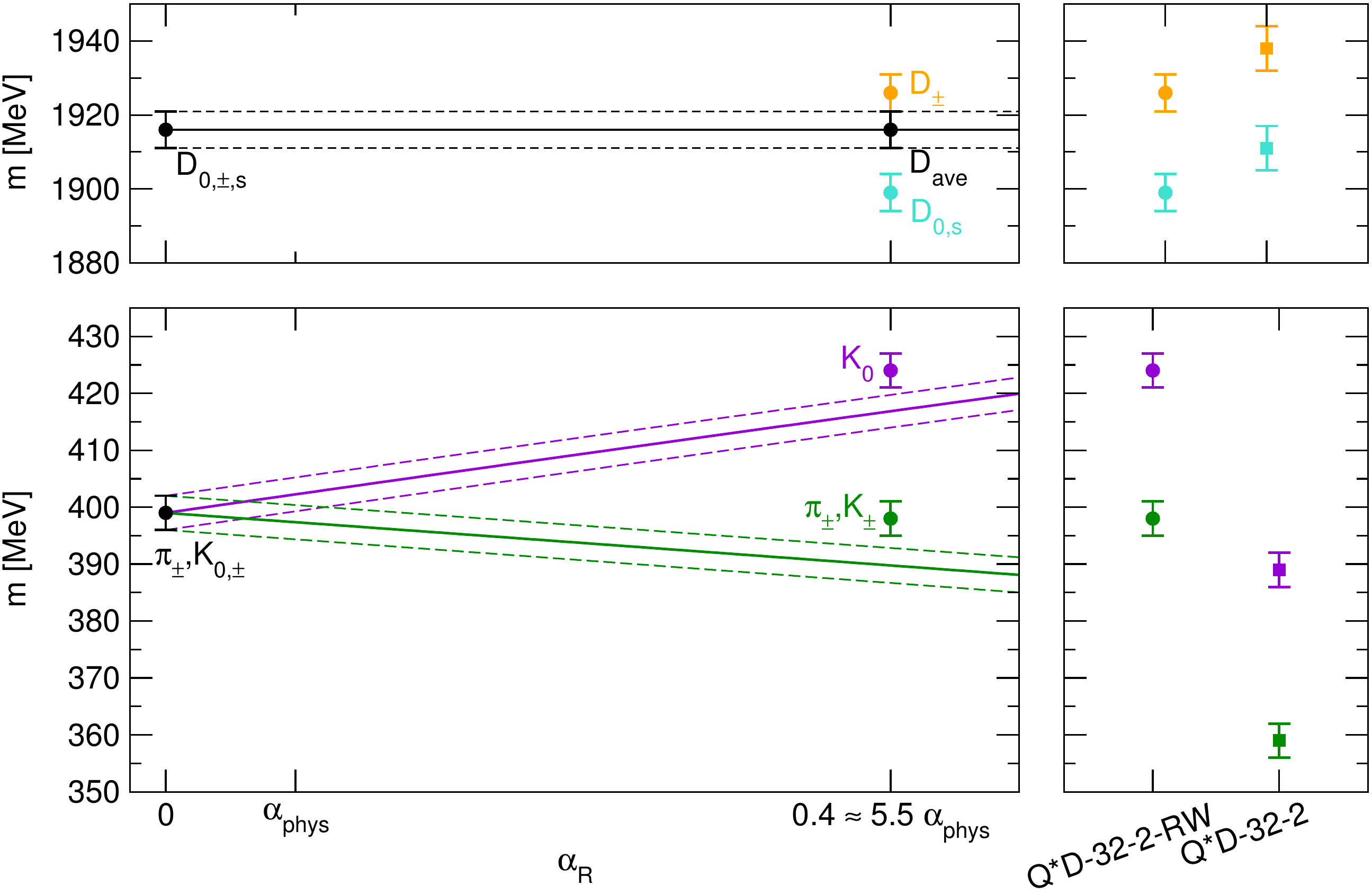}
    \caption{Lines of constant physics. The black points on the left correspond to the ensemble \texttt{QCD-32-1}. The $\pi$ and the $K$ mass that we measured served as starting points to set our lines of constant physics. Following the lines on the right one can see the masses measured on the \texttt{Q*D-32-2-RW} ensemble. While the average of the D-mesons is perfectly on the line of constant physics, the kaons are slightly too heavy. The plot on the right visualizes the shift that we managed to achieve with the reweighting.}
    \label{fig:tuning}
\end{figure}

The most important parameters and observables for our ensembles have been summarized in tables~\ref{tab:summary}, \ref{tab:observables1}, \ref{tab:observables2}. In these tables we include also the run \texttt{Q*D-32-2+RW} which is obtained by reweighting the \texttt{Q*D-32-2} ensemble in the bare quark masses (chosen in such a way to hit the target tuning point). The values of the $\phi_i$ that we measured can be seen in table \ref{tab:observables2}. The resulting lines of constant physics can be seen in figure \ref{fig:tuning}.

\begin{table}[h]

   \bigskip
   
   \begin{center}
      \begin{tabular}{ccccccc}
         \hline
         ensemble & $\alpha$ &
         $\kappa_u$ & $\kappa_d=\kappa_s$ & $\kappa_c$ &
         $c_\text{sw,SU(3)}$ & $c_\text{sw,U(1)}$ \\
         \hline
         \hline
         \texttt{QCD-32-1}    &    0 & 0.13440733 & 0.13440733 & 0.12784 & 2.18859 & 0 \\
         \hline
         \texttt{Q*D-32-1}    & 0.05 & 0.135479 & 0.134524 & 0.12965 & 2.18859 & 1 \\
         \texttt{Q*D-32-2}    & 0.05 & 0.135560 & 0.134617 & 0.129583 & 2.18859 & 1 \\
         \texttt{Q*D-32-2+RW} & 0.05 & 0.1355368 & 0.134596 & 0.12959326 & 2.18859 & 1 \\
         \hline
      \end{tabular}
   \end{center}
   \vspace{-5mm}
   \caption{Simulation parameters. For the first three ensembles, the hopping parameters $\kappa_{u,d,s,c}$ are the
   ones actually used to generate the configurations. For
   \texttt{Q*D-32-2+RW}, the values of $\kappa_{u,d,s,c}$ are the ones used in
   the reweighting procedure.
   }\label{tab:parameters}
\end{table}

\begin{table}[h]

   \bigskip
   
   \begin{center}
      \begin{tabular}{ccccccc}
         \hline
         ensemble & volume & cnfgs & $a$ & $\alpha_\text{R}$ & $L$ & $m_{\pi^\pm}L$ \\
         \hline
         \hline
         \texttt{QCD-32-1} & $64\times32^3$ & 2000 & 0.0539(3) fm & 0          & 1.73(1) fm  & 3.49(3) \\
         \hline
         \texttt{Q*D-32-1}  & $64\times32^3$ & 1993 & 0.0526(2) fm & 0.04077(6) & 1.682(5) fm  & 4.18(2) \\
         \texttt{Q*D-32-2} & $64\times32^3$  & 2001 & 0.0505(3) fm & 0.04063(6) & 1.62(1) fm & 2.90(3) \\
         \texttt{Q*D-32-2+RW} & $64\times32^3$ & 2001 & 0.0510(2) fm  & 0.0407(1) & 1.631(6) fm  & 3.24(3) \\
         \hline
      \end{tabular}
   \end{center}
   \vspace{-5mm}
   \caption{
   \textit{cnfg} stands for the number of
   thermalized configurations for the first three ensembles, or number of
   reweighted configurations for \texttt{Q*D-32-2+RW}. The lattice spacing $a$ is
   calculated by assuming $\sqrt{8t_0} = 0.415 \text{ fm}$ with no error. $L$ is the linear size of the spatial box. The results are preliminary.
   }\label{tab:summary}
\end{table}

\begin{table}[h]

   \bigskip

   \begin{center}
      \begin{tabular}{cccccc}
         \hline
         ensemble &
         $m_{\pi^\pm} = m_{K^\pm}$ & $m_{K^0}-m_{K^\pm}$ & $m_{D^0} = m_{D_s}$ & $m_{D^\pm}-m_{D^0}$ & $\pi \sqrt{3} L^{-1}$ \\
         \hline
         \hline
         \texttt{QCD-32-1}    & 399(3) MeV  & 0 MeV       & 1916(5) MeV & 0 MeV     & --- \\
         \hline
         \texttt{Q*D-32-1}    & 495(3) MeV  & 23.3(5) MeV & 1871(6) MeV  & 32(1) MeV & 639(2) MeV \\
         \texttt{Q*D-32-2}    & 359(3) MeV  & 30(1) MeV & 1911(6) MeV  & 26(2) MeV & 664(4) MeV \\
         \texttt{Q*D-32-2+RW} & 398(3) MeV  & 26(1) MeV & 1899(5) MeV  & 27(2) MeV & 658(3) MeV \\
         \hline
      \end{tabular}
   \end{center}
   \vspace{-5mm}
   \caption{Summary of masses. The masses for charged hadrons have been
   corrected for the universal $O(\alpha_\text{R})$ finite-volume corrections. The
   quantity $\pi \sqrt{3} L^{-1}$ is the smallest energy of a free photon in the
   considered finite box with C$^\star$ boundary conditions in all directions. The results are preliminary.
   }\label{tab:observables1}
\end{table}

\begin{table}[h!]

   \bigskip
   
   \begin{center}
      \begin{tabular}{cccc}
         \hline
         ensemble & $\phi_1$ & $\phi_2$ & $\phi_3$ \\
         \hline
         \hline
         \texttt{QCD-32-1}    & 2.11(3) & --- & 12.09(3) \\
         \hline
         \texttt{Q*D-32-1}    & 3.36(4) & 2.56(5) & 11.93(4) \\
         \texttt{Q*D-32-2}    & 1.81(3) & 2.4(1) & 12.16(5) \\
         \texttt{Q*D-32-2+RW} & 2.20(3) & 2.32(8) & 12.09(3) \\
         \hline
      \end{tabular}
   \end{center}
   \vspace{-5mm}
   \caption{Summary of tuning observables. All ensembles are at the U-symmetric
   point, i.e. $m_d=m_s$ or $\phi_0=0$. The $\phi_{0,1,2,3}$ are described in
   the main text. Our main goal was to tune the
   QCD+QED parameters in such a way that $\phi_{1,3}$ are equal to the QCD runs,
   while $\phi_2 = \phi_2^\text{phys} \simeq 2.37$. The results are preliminary.
   }\label{tab:observables2}
\end{table}


\section{Sign of the Pfaffian}
\label{sec:sign}

Given a quark field $\psi$, we introduce the corresponding antiquark field $\psi^{\mathcal{C}}=C^{-1}\bar{\psi}^T$, where the charge-conjugation matrix $C$ can be chosen to be $i\gamma_0\gamma_2$ in the chiral basis. C$^\star$ boundary conditions for the fermion fields can be written as
\begin{align}
    \begin{pmatrix} \psi(x+L\hat{k})\\
    \psi^\mathcal{C}(x+L\hat{k})
    \end{pmatrix}
    &=
    \begin{pmatrix} \psi^\mathcal{C}(x)\\
    \psi(x)
    \end{pmatrix}
    \equiv
    T\begin{pmatrix} \psi(x)\\
    \psi^\mathcal{C}(x)
    \end{pmatrix}
     \ .
\end{align}
With C$^\star$ boundary conditions the Dirac operator $D$ acts on the quark-antiquark doublet in a non-diagonal way, and it is therefore a $24 V \times 24 V$ matrix. The integration of a quark field in the path integral yields the Pfaffian $\text{pf} \, [CTD]$ in place of the standard fermionic determinant. We rewrite the Pfaffian as
\begin{align}
    \text{pf} \, [CTD]
    &=
    W_{\mathrm{sgn}}
    \left| \text{pf} \, [CTD] \right|
    =
    W_{\mathrm{sgn}}
    \left| \text{det} [D] \right|^{1/2}
    \ ,
    \label{eq:power1/4}
\end{align}
where we have used the algebraic relation $\text{pf} \, [M]^2 = \det [M]$ for a general antisymmetric matrix $M$. In practice we treat the sign $W_{\mathrm{sgn}}$ of the Pfaffian as a reweighing factor. In previous work we have left this out because close to the continuum one expects $W_{\mathrm{sgn}} \simeq 1$.

In order to calculate its sign, it is convenient to relate the Pfaffian to the spectrum of the hermitian Dirac operator $Q=\gamma_5 D$. We first observe that the spectrum of $Q$ is doubly degenerate: if $v$ is an eigenvector of $Q$, then one easily checks that $CT\gamma_5 v^*$ is also an eigenvector of $Q$ with same eigenvalue, and the two eigenvectors are orthogonal. Let $\lambda_{n=1,\dots,12V} \in \mathbb{R}$ be the list of eigenvalues of $Q$, each of them appearing a number of times equal to half their degeneracy. Then one proves that
\begin{align}
    \text{pf} \, [CTD]^2
    =
    \det \left[ D \right]
    =
    \det \left[ Q \right]
    &=
    \prod_{n=1}^{12V} \lambda_n^2
    \ , \qquad\qquad
    \text{pf} \, [CTD]
    =
    \prod_{n=1}^{12V} \lambda_n
    \ .
\end{align}
While the first relation is trivial, the second relation follows from the fact that both sides of the equation are analytic functions of the bare mass $m_0$, and they diverge to $+\infty$ in the $m_0 \to +\infty$ limit. It follows that the Pfaffian is positive (resp. negative) if the number of negative eigenvalues $\lambda_n$ is even (resp. odd). In practice we calculate the sign by following the eigenvalue flow as a function of $m_0$. At very large mass, $Q$ is approximately equal to $m_0 \gamma_5$ and the number of negative eigenvalues is even. As $m_0$ is decreased towards its target value, the Pfaffian flips sign every time an eigenvalue of $Q$ crosses zero. In practice, we follow the flow in the opposite direction, increasing $m_0$ until a crossing becomes unlikely.

Our method is based on two steps: (A) a first fast algorithm identifies a small subset of configurations for which a potential crossing may occur, (B) on these configurations we apply the methods described in~\cite{Campos:1999du,Mohler:2020txx} to determine whether a crossing actually occurs.\footnote{An alternative method has been proposed in~\cite{Bergner:2011zp}.} We describe here only the step A, which is the truly novel ingredient in our calculation.

\begin{figure}
    \centering
        \includegraphics[width=.9\textwidth]{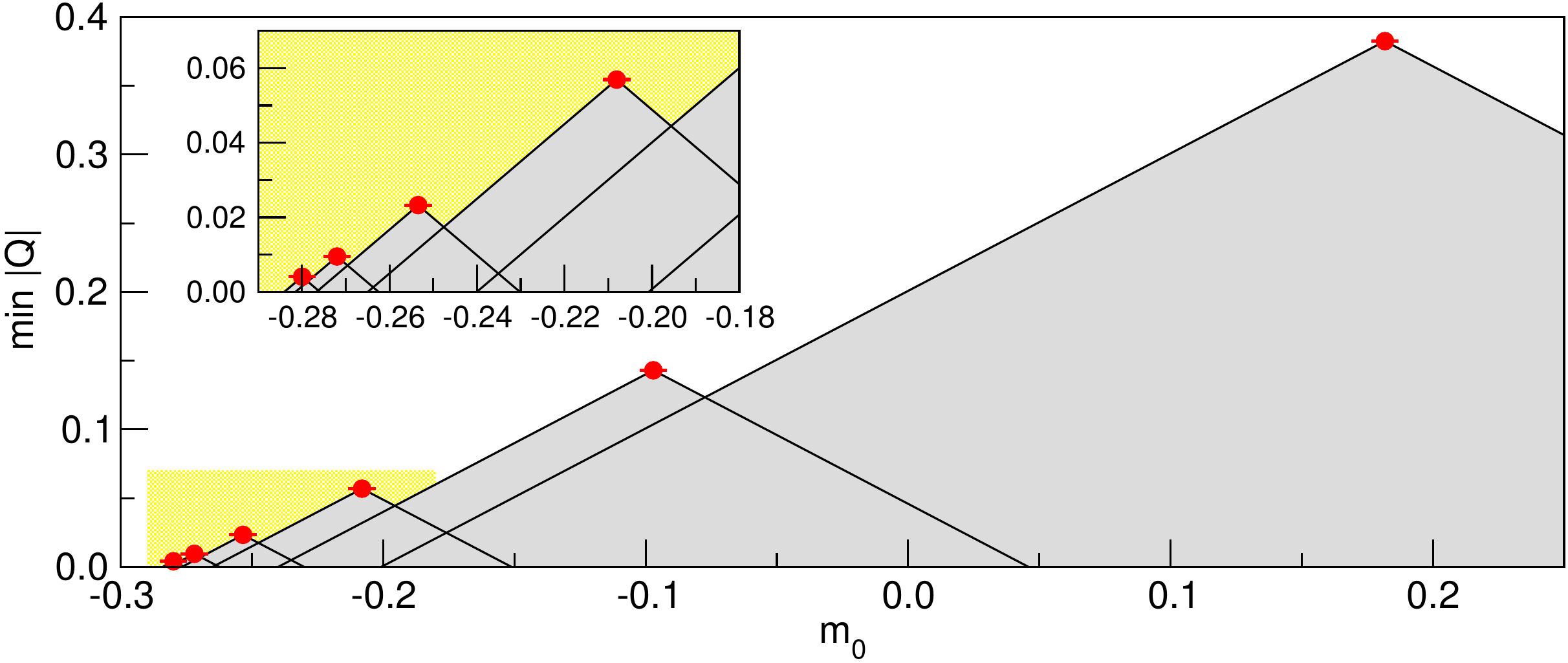}
        \caption{Smallest eigenvalue of $|Q|$ as a function of the valence mass $m_0$ (red points), calculated on a representative configuration (\texttt{QCD-32-1} ensemble). Using the bound on the derivative~\eqref{eq:derivativebound}, one proves that no eigenvalue of $|Q|$ can flow in the grey areas. In particular, no eigenvalue crosses zero in the explored range of masses. One can see that we are able to flow the eigenvalue across two orders of magnitude in only six steps. As evident from the plot an even more efficient version of the algorithm would skip every other step. The inset is a zoom-in of the yellow area.}
        \label{fig:mscan}
\end{figure}

Let $\bar{\lambda}$ be the smallest eigenvalue of the operator $|Q|$, i.e.
\begin{gather}
\bar{\lambda} = \min_n |\lambda_n| \ .
\end{gather}
Since $Q'(m_0) = \gamma_5$, using the Feynman-Hellmann theorem one proves that the derivative of \textit{every} eigenvalue of $Q(m_0)$ satisfies the hard bound
\begin{align}
    \left|
    \lambda'_n(m_0)
    \right|
    &=
    \left|
    \left( 
    \psi_n, Q'(m_0) \psi_n 
    \right)
    \right|
    =
    \left|
    \left( 
    \psi_n, \gamma_5 \psi_n 
    \right)
    \right|
    \le 1 \ .
    \label{eq:derivativebound}
\end{align}
It easily follows that, if $\bar{\lambda}(m_0) > 0$, then no eigenvalue of $Q(\tilde{m}_0)$ crosses zero for $m_0 - \bar{\lambda}(m_0) < \tilde{m}_0 < m_0 + \bar{\lambda}(m_0)$. This observation allows to design the following algorithm, to be run on each configuration:
\begin{enumerate}[noitemsep]
\item Set $m^{(0)}_0 = m_0$ and $n=0$.
\item Calculate $\bar{\lambda}^{(n)} = \bar{\lambda}(m_0^{(n)})$.
\item If $n\ge 1$ and $\bar{\lambda}^{(n)} < \bar{\lambda}^{(n-1)}$ then stop the algorithm and apply step B.
\item If $m^{(n)}_0 > m_0^{\text{max}}$ then stop the algorithm and set $W_{\mathrm{sgn}}=1$.
\item Define $m^{(n+1)}_0 = m_0^{(n)} + c \bar{\lambda}^{(n)}$ and repeat from point 2 with $n \leftarrow n+1$.
\end{enumerate}
The number $c$ could be 2 if we were able to calculate the eigenvalue with infinite precision, but it is chosen to be slightly smaller than 2 for safety. The scan in mass terminates either when the eigenvalue decreases or when the arbitrarily chosen maximal mass $m_0^{\text{max}}$ is reached. In most configurations the eigenvalue does not decrease, which implies that the increment in $m_0$ increases at every iteration. In a few iterations, one can easily cover a couple of orders of magnitude in the eigenvalue $\bar{\lambda}$, as illustrated in figure \ref{fig:mscan}. We stress that in this step we do not need to track eigenvectors, but only the smallest eigenvalue of $|Q|$, which can be efficiently and reliably calculated by applying the power method plus Chebyshev's acceleration to the operator $Q^{-2}$.

\begin{figure}
    \centering
        \includegraphics[width=.95\textwidth]{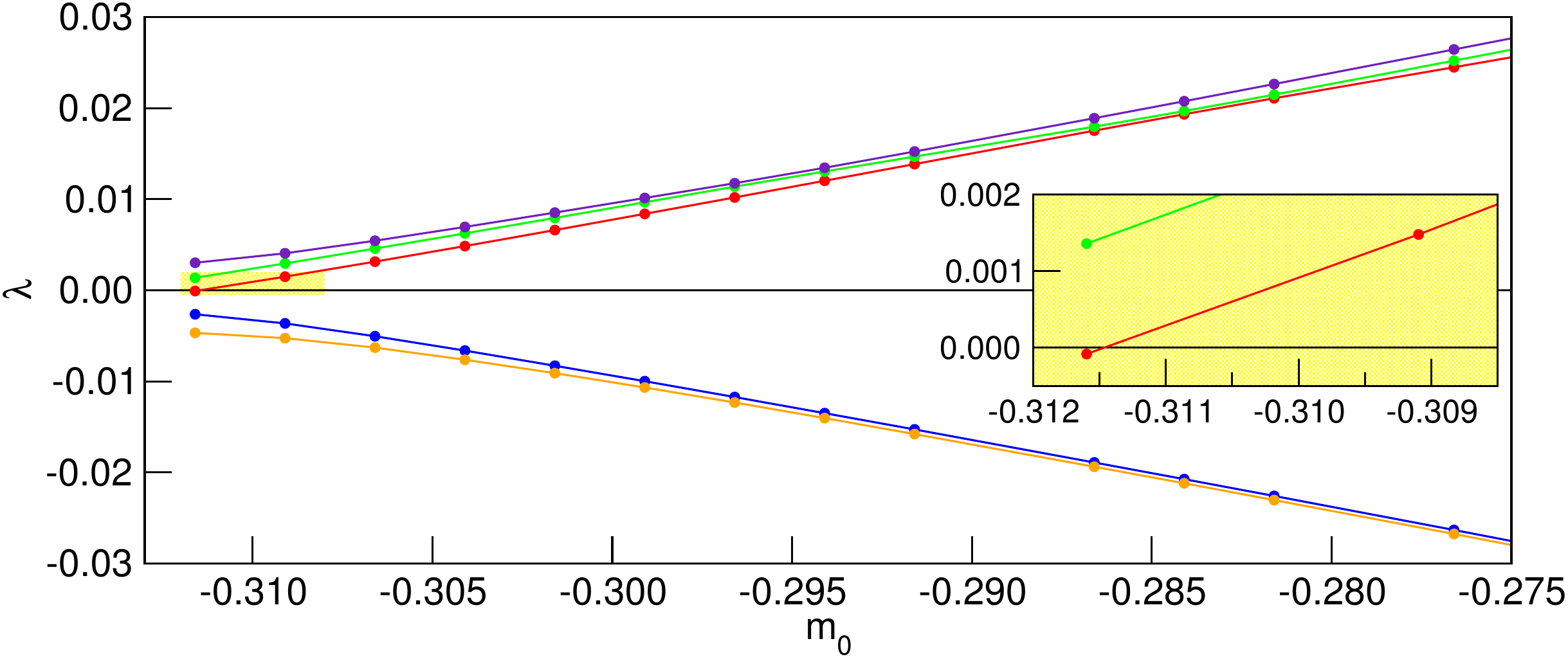}
        \caption{Mass flow for the up quark. We see that for larger bare masses $m_0$ the gap of the Dirac operator $Q$ increases. However between the first and the second measurement the red eigenvalue flows across zero and induces a sign flip of the Pfaffian. This sign was observed on a configuration before the Markov chain thermalized. Hence in the final analysis it did not enter. The inset is a zoom-in of the yellow area.}
        \label{fig:mflow}
\end{figure}

None of the generated ensembles showed a negative sign after the Markov chain thermalized. During thermalization however, some gauge field configurations were present with a negative Pfaffian. One such example from a QCD+QED ensemble can be seen in figure \ref{fig:mflow}.

\section{Reweighting of the mass}
\label{sec:mrw}
In figure \ref{fig:tuning} one can see that, for the \texttt{Q*D-32-2} ensemble, both kaons are roughly 40 MeV too light and the average of the D-mesons is around 10 MeV too heavy. To correct these mistunings we use a reweighting in the mass \cite{Hasenfratz:2008fg}.\footnote{An algorithm for one-flavour mass reweighting has been proposed in~\cite{Finkenrath:2013soa}, however this algorithm does not apply to the case of C$^\star$ boundary conditions since the fermionic determinant is replaced by the Pfaffian.}
We represent the absolute value of the fermionic Pfaffian as in
\begin{align}
    \left|
    \text{pf} \, [CTD_m]
    \right|
    =
    \left|
    \det \left[
    D_m
    \right]
    \right|^{1/2}
    &=
    \frac{
    \left|
    \det\left[
    D_m^{\mathrm{oo}}
    \right]
    \right|
    ^{1/2}
    }{
    \det \left[
    \hat{Q}_m^2
    \right]^{-1/4}
    }
    \to
    \frac{
    \left|
    \det\left[
    D_m^{\mathrm{oo}}
    \right]
    \right|
    ^{1/2}
    }{
    \det \left[
    R(\hat{Q}_m^2)
    \right]
    }
    \label{eq:detQ2} \ .
\end{align}
Here $\hat{Q}_m$ is defined as $\gamma_5 \hat{D}_m$ where $\hat{D}_m$ is the even-odd preconditioned Dirac operator, while $D_m^{\mathrm{oo}}$ is the Dirac operator restricted to and projected onto the odd sites. The operator $(\hat{Q}_m^{2})^{-1/4}$ is replaced by a rational approximation $R(\hat{Q}_m^2)$, whose inverse determinant is stochastically estimated by introducing pseudofermion fields in a standard fashion (for more details see~\cite{Campos:2019kgw}). We choose a rational approximation $R$ of order $(n,n)$ of the form
\begin{align}
    R\left(
    \hat{Q}_m^2
    \right)
    &=
    A
    \prod_{i=1}^{n} \frac{\hat{Q}_m^2 + \nu_i^2}{\hat{Q}_m^2 + \mu_i^2} \ .
    \label{eq:ratapprox}
\end{align}
The parameters are chosen such that $R$ is the optimal rational approximation on a given interval $[r_a,r_b]$, in the sense that the uniform relative error is minimized. 

So for a reweighting of the mass one needs two factors\footnote{The two rational approximations in $W_{\mathrm{mass}}$ can be different. This is useful if the spectral range of the approximated operator changes significantly.}
\begin{align}
    W_{\text{mass}}
    =
    \det\left[
    R
    \left(
    \hat{Q}_{m}^2
    \right)
    R^{-1}
    \left(
    \hat{Q}_{m'}^2
    \right)
    \right],
    &&
    W_{\mathrm{eo}}
    =
    \left|
    \det\left[
    D_{m'}^{\mathrm{oo}}
    (D_m^{\mathrm{oo}})^{-1}
    \right]
    \right|
    ^{1/2}
    \ .
\end{align}
The $W_{\mathrm{eo}}$ factor is only present with even-odd preconditioning and is calculated exactly. The factor $W_{\text{mass}}$ can be written as a product of determinants of positive hermitian operators
\begin{align}
    W_\text{mass} 
    &=
    \prod_{j=1}^{2n}
    \det\left[
    \left(
    1
    +
    \delta \hat{D}
    S_{j}
    \right)^{\dagger}
    \left(
    1
    +
    \delta \hat{D}
    S_{j}
    \right)
    \right] \ .
    \label{eq:Wmass}
\end{align}
The difference of the Dirac operators $\delta\hat{D}=\hat{D}_{m'}-\hat{D}_m$ can be worked out analytically. The operator $S_{j}$ is defined as
\begin{align}
    S_{j}
    &=
    \left(
    \hat{D}_{m} + i \gamma_5 \nu_{j}
    \right)^{-1}
    -
    \left(
    \hat{D}_{m'} + i \gamma_5 \mu_{j}
    \right)^{-1}
    -
    \left(
    \hat{D}_{m'} + i \gamma_5 \mu_{j}
    \right)^{-1}
    \delta \hat{D}
    \left(
    \hat{D}_{m} + i \gamma_5 \nu_{j}
    \right)^{-1} \ .
\end{align}
In practice, every determinant from eq. \eqref{eq:Wmass} is estimated stochastically, i.e.
\begin{align}
    W_\text{mass} 
    &=
    \prod_{j=1}^{2n}
    \det\left[
    1
    +
    R_{j}
    \right]
    =
    \prod_{j=1}^{2n}
    \left(
        \frac{1}{N_j}
        \sum_{l=1}^{N_j}
        e^{-\left(\eta_{jl},R_j \eta_{jl} \right)}
    \right) \ .
\end{align}
The hermitian operator $R_{j}$ is defined as
\begin{align}
    R_{j}
    &=
    \left(
    \delta \hat{D}
    S_{j}
    \right)^{\dagger}
    +
    \left(
    \delta \hat{D}
    S_{j}
    \right)
    +
    \left(
    \delta \hat{D}
    S_{j}
    \right)^{\dagger}
    \left(
    \delta \hat{D}
    S_{j}
    \right) \ ,
\end{align}
and the complex stochastic sources $\eta$ have support on the even lattice sites and a probability distribution proportional to $e^{-(\eta,\eta)}$.

We computed the reweighting factor with a single stochastic source for every factor. Investigating the effects of the reweighting shows that in our case the mistuning was small enough, so that we do not observe an increase in the errors. In the tables \ref{tab:observables1} and \ref{tab:observables2} this can be seen explicitly. The reweighting induces a slight shift in the lattice spacing, but within errors the electromagnetic coupling stayed the same. From table \ref{tab:summary} one can see that only $\phi_1$ is off the line of constant physics after reweighting.


\section{Summary}
For the first time we have computed the sign of the Pfaffian and included it into our analysis. Thus we are simulating the full path integral. We have presented a two-part algorithm that can efficiently detect gauge field configurations that give a negative fermionic Pfaffian. For our ensembles we did not observe any configurations with negative sign once the Markov chain thermalized. It will be interesting to see at which pion masses and what lattice spacings negative signs actually become a problem.

After the reweighting in the mass, no significant increase in the error of any observable was observed. For a larger shift and larger volumes we expect an increase in the errors. A preliminary analysis showed that the computation of the mass reweighting factor is about $40\%$ cheaper than the generation of a new ensemble without considering thermalization. Since the tuning of the parameters in a fully dynamical QCD+QED simulation is a complex task and in practice requires the generation of several tuning ensembles, it is interesting in what regime of quark masses and for what volumes the reweighting gives reasonable results. 


\paragraph{Acknowledgements.}
We would like to thank Daniel Mohler and Stefan Schaefer for sharing their code to calculate the eigenvalue flow with us.
The research of AC, JL and AP is funded by the Deutsche Forschungsgemeinschaft (DFG, German Research Foundation) - Projektnummer 417533893/GRK2575 “Rethinking Quantum Field Theory”.
The work was supported by the North-German Supercomputing Alliance (HLRN) with the project bep00085.
The work was supported by the Poznan Supercomputing and Networking Center (PSNC) through grant numbers 450 and 466. 
The work was supported by CINECA that granted computing resources on the Marconi supercomputer to the LQCD123 INFN theoretical initiative under the CINECA-INFN agreement.
The authors acknowledge access to Piz Daint at the Swiss National Supercomputing Centre, Switzerland under the ETHZ's share with the project IDs go22 and go24.

\setlength{\bibsep}{0pt}

\end{document}